# Effective factors in agile transformation process from change management perspective

Taghi Javdani Gandomani, Hazura Zulzalil, Abdul Azim Abd. Ghani, Abu Bakar Md. Sultan
Faculty of Computer Science and Information Technology
University Putra Malaysia, UPM
Serdang, Malaysia
tjavdani@yahoo.com, {hazura, azim, abakar}@fsktm.upm.edu.my

*Abstract*:

**After introducing agile approach in 2001, several agile methods were founded over the last decade. Agile values such as customer collaboration, embracing changes, iteration and frequent delivery, continuous integration, etc. motivate all software stakeholders to use these methods in their projects. The main issue is that for using these methods instead of traditional methods in software development, companies should change their approach from traditional to agile. This change is a fundamental and critical mutation. Several studies have been done for investigating of barriers, challenges and issues in agile movement process and also in how to use agile methods in companies. The main issue is altering attitude from traditional to agile approach. We believe that before managing agile transformation process, its related factors should be studied in deep. This study focuses on different dimensions of changing approach to agile from change management perspective. These factors are how to being agile, method selection and awareness of challenges and issues. These fundamental factors encompass many items for agile movement and adoption process. However these factors may change in different organization, but they should be studied in deep before any action plan for designing a change strategy. The main contribution of this paper is introducing and these factors and discuss on them deeply.**

*Keywords-Agile software development, Agile transformation, Change management strategy, Agile methods, Agile adoption*

I. INTRODUCTION

Agile approach as a reaction [1] to traditional approach in software development was introduced by agile manifesto [2]. In this manifesto new values were considered in software industry and also several principles were introduced as agile underpinning in the organization [2]. In competitive world, technology and industry grow too fast and so, requirements change rapidly and agile methods can support these changes effectively [3]. Agile values have attracted many companies to change their production approach from plan-based methods to agile methods. Several well-known companies have migrated to agile and now are using these methods even in some of their projects [4-6]. Several studies have done in how using agile methods and also several case studies were reported in agile migration. A critical issue for using agile methods is that companies and organization should change their approach fundamentally and this is not an easy process[7]. Several studies are conducted in transforming to agile from different views. They are based on specific method [8], sepecific culture [9] or specific organization [10, 11]. However there are some guidelines or basic framework offered by some studies for handling migration process [12, 13] so far, but still it is need to study more in deep and from various perspective [13],it means that agile migration is still a hot research area in software engineering. Since agility affects all aspects of organization, agile migration should be studied in a wider perspective. For first step, effective factors for this organizational mutation should be explored from substantive data in industry. Based on our literature, there are many factors that should be considered in moving to agile, but in a wide perspective and from change management strategy perspective, we have classified them in three main areas: how to being agile, method selection and awareness of challenges.

The next sections of this paper are organised as follow: in Section 2 we show that how an agile method can be fixed in company, in Section 3 we discuss about th role of agile methods in transformation process, in Section 4 we investigate the challenges and issues in agile transformation and in Section 5 we discuss on role of these factors in change management strategy and finally in section 6 we present conclusion and our future work.

## II. CHOICES FOR BEING AGILE

Companies and organizations based on their needs and limitations should decide about how being agile. Indeed they are able to choose only some agile activities or using agile for only some steps of their software product line or become agile completely even by using more than one agile method. The main options for this decision are explained in the next sections.

### A. Tailoring: using agile practices beside plan-driven methods

In this approach companies are not interested in fundamental change in their process, but they want to use agile activities and practices in only some specific stages with reasonable change only. This approach was the first choice for companies that have been relied on CMMI quality model. They needed to maintain their quality level in CMMI and if it was possible to take advantages of agility [14-16]. However, there are some reports on successful agile adoption in CMMI companies [17], but some of CMMI practice areas are in conflict with agile approach [18, 19]. It should be noted that both of these approach have their own benefits, but agile provide new values. Tailoring is a good choice especially for those companies which their customers ask them a rigid and disciplined development methodology. By tailoring agile practices in their disciplined process, they can provide some agile values simultaneously with meeting customer's requirement.

### B. Localization: using agile methods by some modification

In this approach despite of previous approach companies accept essential and fundamental change in their organizations. The main issue is that because of some limitations they are not able to adapt with all agile activities; so, they should customize some of the agile activities or ignore them. Sometimes this approach is the only option for fulfillment organization, project or management requirement [20]. This approach is also beneficial in early stages of agile transformation or when stakeholders are not experienced. In these cases, it is necessary that some of agile practices like group decision making to ignore [21] . Furthermore, sometimes customer collaboration is not possible and so agile customer related activities should be done in traditional way [22]. Project limitation and incompatibility of agile with pilot project also forces companies to use agile methods in customized versions [23]. It seems that companies choose this option only because of necessary constraints in their organizations and projects.

*C. Adpotion: embrassing agile completely*

In this approach like localization, companies accept essential change in their organizations. In this option, managers try to overcome internal and external constraints to meet maximum agile values. Of course there are a lot of obstacles and constraints that should be identified before migration to agile. There are many studies on obstacles and challenges in agile adoption [24, 25]. Also a few researches have been done for proposing guidelines or frameworks to facilitate agile movement [12]. Furthermore, several case studies have been reported about journey of agile movement in different companies [26-28]. Agile adoption is the best way for achieving maximum agile values and this is a fertilize area for researches mainly because this process should be studied from different perspectives.

### III. METHOD SELECTION

There are several agile methods that each of them has its specific characteristics and activities. Although all of the agile methods are founded based on agile values but each of them emphasizes of some values more than others. Cohen et al. have explained more popular agile methods in their study [29]. One of the critical and important issues in agile transformation is method selection. Indeed for finding best suitable method and facilitate the movement process, a comprehensive study about abilities or disabilities of each method should be done.

Generally agile methods can be divided in two main groups based on their fundamental practices: software development and software management. In other words, some of them mainly focus on the managing of software projects and others on software development process. However a combination of both group is more useful in almost all companies, but some companies choose only one method from one of the mentioned group. There are some valuable studies in comparing agile methods and discussion on capabilities of them. In one of them, differences of agile methods is studied in a comparative analysis research from various perspective [30]. In another one, implications and applicability of different methods in industry is studied [31]. Also some other studies compare two specific agile methods in deep from various views [32, 33].

Managers should consider their goals, needs, organization capabilities and constraints in choosing appropriate methods for their projects. Such decision can affect the future of their companies. Wrong decision in method choosing strongly will affect on success of agile migration. However, some studies have focused on decision making in method selection [34, 35], but they are not enough. This issue should be studied as a significant part or agile change management strategy.

### IV. AWARENESS OF CHALLENGES AND OBSTACLES

For agile transformation, all aspects of organization should change and this fundamental change cause many problems and challenges. Agile change management strategy should be prepared only after recognition of challenges, obstacles and barriers. Previous studies show that different challenge might be seen in this process [24]. Some of them are related in management and organization. For instance, changing attitude from "command and control" to "leadership and collaboration" is a big issue [36]. Coaching and mentoring in this process is difficult, because not only technical problems should be solved but also mindset of peoples should be considered. Knowledge management is another issue in agile methods. While in plan-driven methods heavy documentation and rigid reports are required, in agile methods knowledge is tacit and in the head of the stakeholders and act as a barriers from perspective of traditional senior managers [37]. In process context, changing traditional life cycle to iterative and evolutionary model is a big issue. It is mainly because of effects of the process model on different parts of organization [28]. Also different measurement practices is another issue in this domain [38]. A lot of obstacles are reported in human aspects [39]. Sometimes people cannot forget their previous role and resist against the change [40]. For instance role of project manager is a challenge in this process [41], because they should be leadership instead of commander. In multi-sites and international organizations lack of face-to-face communication, co-located working, different cultures and time zone offset also reported as major obstacles [42, 43]. Indeed in such companies the big issue is communication which is a principle in agile methods. Discussion in this area is long and beyond the scope of this paper and could be done in a detail discussion.

## V. ROLE OF THE FACTORS IN CHANGE MANAGEMENT STRATEGY

Change management strategy should manage the process of agile transformation. Since this process is pervasive, all effective factors should be studied in it. The mentioned factors are the most important factors in agile transformation and any shortcoming about each of them causes many problems for migration process. "Fig. 1" shows that Change management strategy is surrounded by these factors. In one hand, managers should be aware about challenges and obstacles and in the other hand they should choose the most suitable agile method(s) for their projects based on their requirements and constraints. Also they should decide that how they want to be agile. They should find best answer for how being agile and which methods are the best for them. Answer of this questions are underpinning of change management strategy.

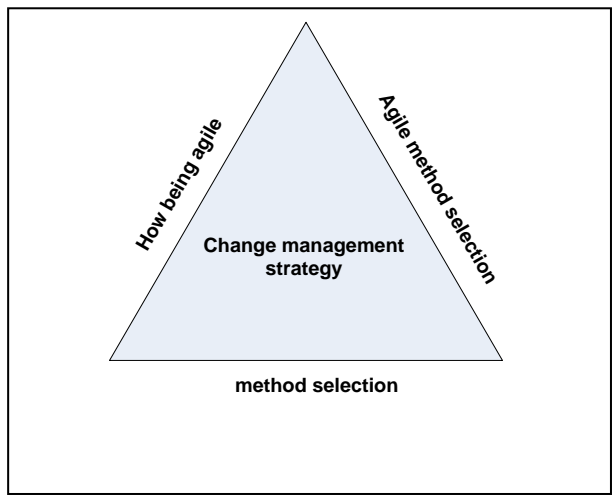

Figure 1. Change management effective factors

## VI. CONCLUSION AND FUTURE WORK

Changing software development approach from traditional to agile is a fundamental change in organization and should be managed via comprehensive change management strategy. Agile transformation should be considered as a disciplined process and before inception should be known completely. Main aim of this paper is investigation on most effective factors that influence agile transformation. We classified these factors in three main groups: how to being agile, method selection and awareness of obstacles and barriers. Each of these factors provides many items that should be considered in designing change management strategy. Our future work is suggestion of a change management strategy for agile transformation. This strategy should be emerged from substantive experiments and not based on the conceptual practices.